# Identification of information tonality based on Bayesian approach and neural networks

D.V. Lande

ElVisti Information center, Kiev, Ukraine

*A model of the identification of information tonality, based on Bayesian approach and neural networks was described. In the context of this paper tonality means positive or negative tone of both the whole information and its parts which are related to particular concepts. The method, its application is presented in the paper, is based on statistic regularities connected with the presence of definite lexemes in the texts. A distinctive feature of the method is its simplicity and versatility. At present ideologically similar approaches are widely used to control spam.*

**Tonality issue**

The pace of development, dynamics and bodies of information space of the Internet transform it into information flow [1]. Studying the flow of news information, published on the pages of Web-sites, should use an absolutely new mechanisms, as the conventional methods cannot always cover even the representative part of this flow (nothing to say about the whole flow). A traditional expert estimation of text information appears to be inefficient for super large and super dynamic text files. One of the aspects of the analysis of text information from current information flows, namely, the estimation of text tonality, is considered in this paper. In the context of this paper text tonality refers to positive, negative or neutral emotional tone of both the whole information and its separate parts, related to particular concepts such as persons, organizations, brands etc.

The system "VAAL" [2] is the best known in the sphere of automation of the process of text tonality identification; it is oriented on an emotional-ligical analysis. This system is based on making a frequency word list, its analysis as to the availability of particular words which allow, with some probability, to determine psych-linquistic categories

Another approach, oriented on the use of linguistic algorithms, statistic text analysis, which takes into consideration modal characteristics of the situation, modus meanings and author's attitude to a described situation, is a promising one; however, it is quite resource-consuming and not versatile enough.

Besides, the nature of the majority of electronic publications makes it possible to estimate text tonality, its emotional tone directly by its vocabulary, which is close to the approaches used in the system "VAAL".

The method, its application is presented in the paper, is based on statistic regularities connected with the presence of definite lexemes in the texts, naïve Bayesian approach and neural networks (realization of two-layer perceptron). A distinctive feature of the method is its simplicity and versatility, the accuracy of the estimation being regulated parametrically in a rather wide range. At present ideologically similar approaches are widely used to control spam and they yield good results [3, 4].

However, it is necessary to mention that the task of the identification of information tonality is more complicated than that of spam based on the text analysis. To identify spam implies two hypotheses (spam, not spam); when we identify tonality three aspects are involved: emotional positive, negative, neutral tone, and quite frequently there is a necessity to check the combination of these hypotheses (for example, to determine the level of text "expressiveness").

On the other hand, unlike the problem of spam identification where estimation of individual documents can be close to one-valued, in case of tonality identification even various people-experts do not always agree. Here the situation approaches the estimation level of relevancy-pertinency in information search.

**Bayesian approach to the estimation of information tonality**

As the suggested approach is close to the one which is used in Bayesian anti-spam filters, let us first consider the application of Bayesian theorem to solve spam problems. Bayesian method envisages the use of an estimation tool – two bodies of electronic letters: one of them is made of spam, the other one – of usual letters. Word frequency is calculated for each body, then weight estimation is made (from 0 to 1), which characterizes conditional probability that the information with this word is spam. Meanings of the weights, close to ½, are not considered in integrated calculation, and the words with such weights are ignored and removed from dictionaries.

According to the method, suggested by Paul Graham, if information contains $n$ words with weight meanings $w_1...w_n$, the estimation of conditional probability, that a letter is spam, which is based on the data from estimation bodies, is calculated with help of a formula:

$$Spm = \frac{\prod w_i}{\prod w_i + \prod (1 - w_i)}. \tag{1}$$

This formula is explained with the following considerations. It is assumed that $S$ is an event which means that a letter is spam, $A$ is an event which means that a letter has word $t$. Then in accordance with Bayesian formula, it is true:

$$P(S|A) = \frac{P(A|S)P(S)}{P(A|S)P(S) + P(A|\overline{S})P(\overline{S})}. \tag{2}$$

If it is not clear initially whether a letter is spam or not, then based on the experience, it is suggested that $P(\overline{S}) = \lambda P(S)$, and it comes from (2) that:

$$P(S|A) = \frac{P(A|S)}{P(A|S) + \lambda P(A|\overline{S})}. \tag{3}$$

Further, formula (3) is generalized in the following way. It is assumed that $A_1$ and $A_2$ are events which mean that a letter contains words $t_1$ and $t_2$. Another assumption is that these events are not dependent (this is exactly why the method is called "naïve" Bayesian). Conditional probability that the letter, containing both words ($t_1$ and $t_2$), is spam, is equal to:

$$P(S|A_1 \& A_2) = \frac{P(A_1|S)P(A_2|S)}{P(A_1|S)P(A_2|S) + \lambda P(A_1|\overline{S})P(A_2|\overline{S})}$$
$$= \frac{p(t_1)p(t_2)}{p(t_1)p(t_2) + \lambda(1 - p(t_1))(1 - p(t_2))}. \tag{4}$$

Formula (1) is the generalization of formula (4) in case of a randomized number of words and $\lambda=1$. We have to mention that the meaning $\lambda=1$ is the most widely used in anti-spam filters. On the one hand, it simplifies calculations, but on the other hand, it distorts reality and reduces quality performance of these programs considerably.

In practice, based on the dictionaries/lists of words, which are constantly modified, meaning *Spm* is calculated for each message. If it exceeds some threshold level, the information is considered to be spam.

When information tonality is estimated, hypothesis space will contain: $H_{-1}$ – tonality is negative, $H_0$ – tonality is neutral and $H_1$ – tonality is positive, $\overline{H_1}$ - tonality is not positive. Words, which are typical for these documents, are picked from a document body with positive tonality. From them words *t* with meaning $p(t|H_1)$, exceeding ½, for example 0.6, are picked. Such words are called tone-colored or simply tonal, having estimating semantics.

To simplify the model, let us assume that for all chosen terms weight will be the same, equal to α (it can change when teaching the model). Then formula (1) will look like:

$$Spm(x) = \frac{\alpha^x}{\alpha^x + \lambda(1-\alpha)^x}. \qquad (5)$$

where *x* – the number of weighty words from the point of view of tonality in the information message, α – weight.

As it is seen from Fig. 1 (α = 0.6, λ = 1), the availability of 10 words, typical for positive tonality, guarantees that the information message will possess the same property.

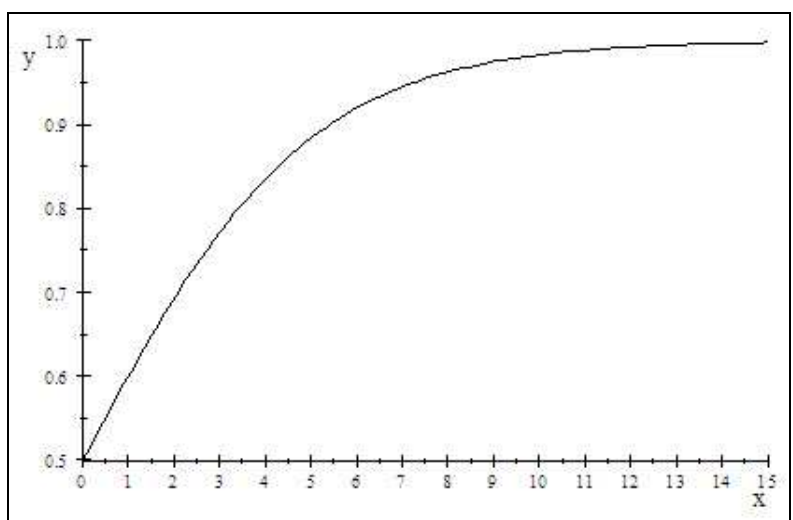

*Fig. 1. A plot of a function Spm(x)*

To estimate a hypothesis about negative tonality of information ($H_{-1}$), a list of words with "negative tonality" and the same formula (5) can be used. Besides, as positive and negative tonalities are somewhat antagonisms, a final decision about information tonality is made taking into consideration a difference of meanings of weight hypothesis estimation $H_1$ and $H_{-1}$. A threshold meaning of this quantity - β is defined in the process of adjusting (teaching) the system.

It is necessary to make one more remark, prompted by practical experience. Negative information tonality in the Internet is almost always expressed more vividly than a positive one. To compare tonalities while calculating the weight of negative tonality, meaning *x* in formula (5) is decreased slightly through multiplying it by empirically defined constant $\gamma \in (0, 1)$.

In some cases the documents, which have high weight meanings of both positive and negative tonality, present certain interest for analysts. The difference of these weights may be minimal, i.e., the document may be characterized as neutral. In addition, it will get a characteristic of "expressive" tonality.

**Model of neural network**

The realization of a given algorithm will be presented in the form of neural network [5] (Fig. 2). The first layer of this network contains two neurons – determinants of weight meanings of positive and negative tonality (positive and negative neurons). We can assume that the number of dentrites of each neuron is equal to that of layers from the dictionary of a natural language. Input signals – meanings $x_1...x_n$, which correspond to input words come to a neuron entry. In this case $x_i=1$, if the word with number $i$ entered, then $x_i=0$. Weight meanings (synapse weights), which correspond to these words, are equal to $w^+_1...w^+_n$ for a positive neuron and $w^-_1...w^-_n$ – for a negative one. It is these weght meanings that can change in the process of teaching perceptron. The adders calculate the meanings of $NET^+$ and $NET^-$, correspondingly. Neuron conductance is calculated with formula (5), the argument is menaing $NET^+$ for a positive neuron and $\gamma NET^-$ for a negative one. Both neurons, via axons/neurites, give gradient meanings, $OUT^+$ and $OUT^-$, which are input signals for a neuron of the second level, whose adder calculates the difference $OUT^+$ and $OUT^-$, and conductance function gives a gradient result as it is shown in Fig. 2.

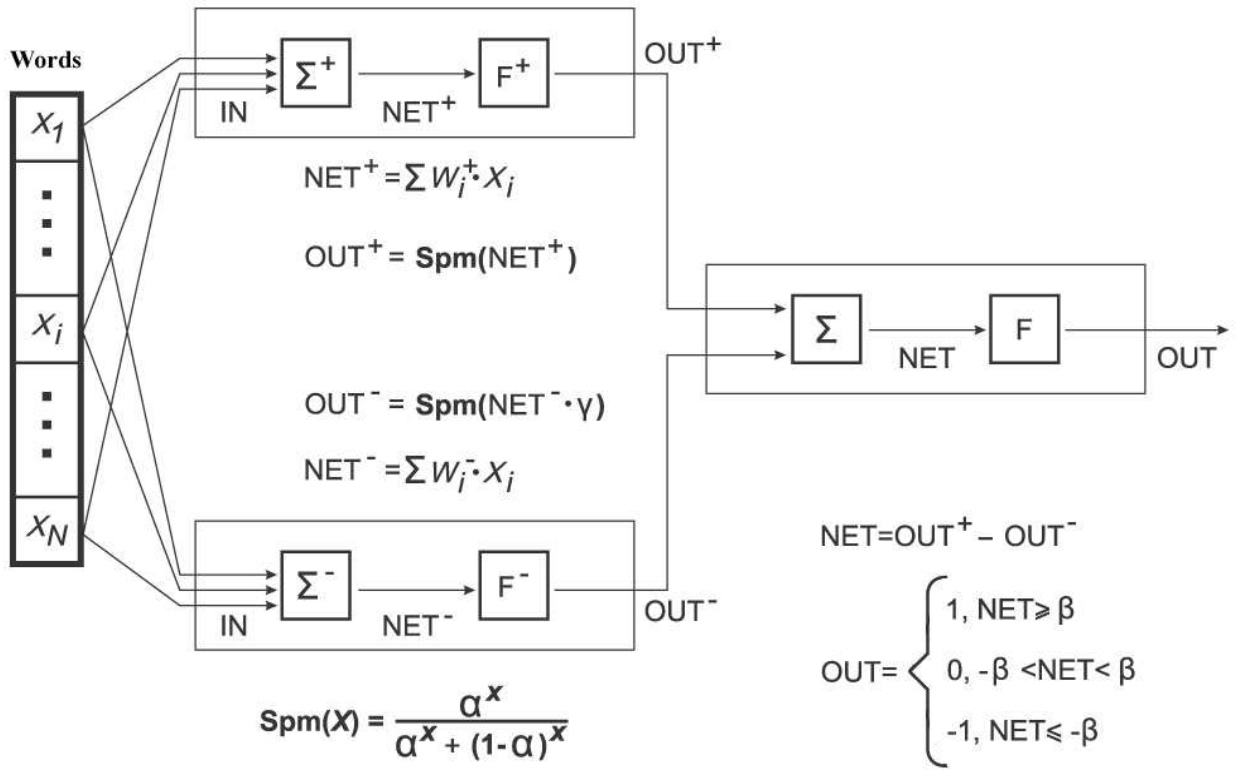

*Fig. 2. A two-layer perceptron which determines text tonality*

**Practical application of the approach**

A suggested model is realized in the content-minitoring InfoStream system, which is used to implement the tasks of gathering news information from open web-sites, its systematization and ensuring the access to it in retrieval/search regimes. At present this system covers 3500 sourecs – over 60000 unique news information within 24 hours. To navigate in these information resources and to specify inquiries, a mechanism of information portrait was worked out; it is a multi-aspect selection of choosing parameters based on the initially made request/inquiry. In the information portrait the meaning of information tonality is used as one of the parameters, the specification of which makes it possible to single out publications of negative or positive tonality, corresponding to the topics, defined by the initially entered inquiry. Another feature of InfoStream system is to monitor the appearance dynamics of the concepts, which correspond to the inquiries entered by the users. The messages, marked with positive or negative tonality, form segments of green or red color on a corresponding diagram.

### From text tonality to concept tonality

A suggested algorithm is used as an instrument to identify general information tonality, while tonality of separate concepts, covered by the message, does not always correspond to the tonality of the whole information. We can assume (according to experts-analysts, such assumption is quite justified in practice) that when a large information channel corresponding to some inquiry is analyzed, an emotional tone of a target concept will coincide with the integral estimation of the information channel.

More accurate results can be received when a fragment, e.g. a paragraph, containing a concept of interest, a sentence or even a part of the sentenece, is estimated rather than the whole information.

It is clear, that such approach does not ensure accuracy in every case, and the information being divided into small parts may lose the wholeness of its meaning. However, we stress again, that for representative information channels relating to a target concept, a suggested methodology appears to be not only "transparent" but also quite efficient due to statistic regularities.